\documentclass[fleqn,usenatbib,useAMS]{mnras}

\usepackage{graphicx}	
\usepackage{amsmath}
\usepackage{amssymb}	
\usepackage{bm}
\usepackage{physics} 
\usepackage{multicol}
\usepackage{array}

\newcommand{\p}{\partial}

\newcommand{\rmk}{\mathrm{k}}

\newcommand{\rmi}{\mathrm{i}}

\newcommand{\be}{\mathbf{e}}

\newcommand{\bF}{\mathbf{F}}
\newcommand{\bk}{\mathbf{k}}
\newcommand{\bu}{\mathbf{u}}
\newcommand{\bv}{\mathbf{v}}
\newcommand{\bx}{\mathbf{x}}

\title[The saturation of VSI]{The saturation of the VSI in protoplanetary disks via parametric instability}

\author[Cui \& Latter]{
Can Cui\thanks{E-mail: \href{mailto:cc795@cam.ac.uk}{cc795@cam.ac.uk}}
and
Henrik N. Latter
\\
DAMTP, University of Cambridge, CMS, Wilberforce Road, Cambridge CB3 0WA, UK
}

\begin{document}
\label{firstpage}
\pagerange{\pageref{firstpage}--\pageref{lastpage}}
\maketitle

\begin{abstract}

The vertical shear instability (VSI) is a robust and potentially important phenomenon in irradiated protoplanetary disks (PPDs), yet the mechanism by which it saturates remains poorly understood. Global simulations suggest that the non-linear evolution of the VSI is dominated by radially propagating inertial wavetrains (called `body modes'), but these are known to be susceptible to a parametric instability. In this paper, we propose that the global VSI saturates via this secondary instability, which initiates a redistribution of energy from the large scales to smaller-scale inertial waves, and finally into a turbulent cascade. We present an analytic theory of the instability in a simple idealised model that captures the main physical and mathematical details of the problem. In addition, we conduct numerical simulations with the SNOOPY code to consolidate the theory. Once the parametric instability prevails, the VSI is likely far more disordered and incoherent than current global simulations suggest. We also argue that it is challenging to capture parametric instability in global simulations unless the radial resolution is very fine, possibly $\sim 300$ grid cells per scale height in radius. 

\end{abstract}


%%%%%%%%%%%%%%%%%%%%%%%%%%%%%%%%%%%%%%%%%%%%%%%%%%

\section{Introduction}\label{in}

Research over the last decade has revealed that the vertical shear instability (VSI) is a robust phenomenon in irradiated protoplanetary disks (PPDs). The magnetorotational instability (MRI; \citealp{c61,bh91}), which prevails in many astrophysical disks, is predicted to be quenched, or at least significantly weakened, in the bulk of a PPD due to its extremely low ionization fraction \citep[e.g.][]{turner_etal14,lesur20}. Instead, the majority of the angular momentum is likely transported by laminar magnetized disk winds \citep{bs13,gressel_etal15,bai17,gressel_etal20,cb21}. On the other hand, purely hydrodynamic instabilities are frequently invoked \citep[e.g.][]{fl19,lu19} in order to explain the observed level of turbulence in the disk \citep{teague_etal16,flaherty_etal17,flaherty_etal18,flaherty_etal20}. The VSI is at the forefront of these hydrodynamic instabilities, because the thermodynamic conditions in PPDs are highly conducive to its onset and it likely extends over a significant portion of the disk \citep{malygin_etal17,pk19,lu19}. 

The VSI is the disk analogue of the Goldreich--Schubert--Fricke instability \citep{gs67,fricke68}, originally discovered in the context of differentially rotating stars. Though first discussed by \citet{ub98}, the importance of the VSI to PPDs was only brought out much later through the global simulations of \citet[][hereafter N13]{nelson_etal13}. Subsequent linear analyses in incompressible, vertically global, and fully global models with and without magnetic fields elucidated its mathematical properties \citep[N13;][]{bl15,ly15,lp18,cl21}. Follow up numerical simulations have investigated the non-linear development of the instability, and added relevant physics such as radiative transfer, dust, and non-ideal magnetohydrodynamics (MHD) \citep[e.g.][]{sk14,flock_etal17,lin19,flock_etal20,cb20,schafer_etal20,ll21}. One important feature of both the VSI's linear theory and non-linear saturation in global models is the importance of remarkably coherent wave patterns. The VSI has a preference to develop radially travelling inertial waves, called body modes \citep[N13;][]{sk14,bl15}.

Although a large body of numerical simulations describing the non-linear behaviour of the VSI has accumulated, an important topic that has received insufficient attention is its saturation mechanism. Presently, there is no secure physical picture of the processes that limit the VSI growth and control its consequent nonlinear quasi-steady state, i.e. the body modes that predominate within it. Such a theory, if developed, would allow us to assess whether current numerical simulations are accurately capturing the VSI physics, especially during the later stage of its non-linear evolution, and whether we can generalise these simulation results to more realistic contexts beyond current numerical capabilities.

We envisage the saturation to be potentially comprised of two distinct processes. The first is the elimination of the vertical shear. This could be accomplished by the large-scale vertical motions of the VSI, leading to the redistribution of angular velocity and heat. In the limit of weak or no external thermodynamic forcing, this erasure of shear is the dominant saturation process, and the VSI will ultimately die out for want of a free energy source. In general, however, the radiative forcing from the central star and the VSI's vertical transport will come into a quasi-steady balance, which will yield some degree of vertical shear and hence a finite VSI amplitude. This brings us to the second saturation process: the energy transfer from the large-scale VSI modes to the small scales, where it is dissipated. In the limit of very strong thermal driving (e.g. in locally isothermal simulations), it is this process that will control the saturation of the VSI.

In this paper, we focus on the second ingredient in the saturation process: how the energy continuously fed into the large-scale body modes is disseminated to shorter-scale modes. We find that this direct cascade of energy can be started by parametric instability, where the dominant body modes fall into resonance with pairs of inertial waves \citep{bordes12}. We expect each pair of secondary modes to subsequently interact with smaller-scale inertial waves, and instigate a wave-turbulent cascade (e.g. \citealp{ns11}). The resulting energy flux, when balanced against the rate of energy input by the VSI body modes, should set the saturated turbulent level.  

To understand how the parametric instability instigates the direct cascade of energy, we calculate the linear stability of radially propagating inertial waves (body modes), and also simulate their development with the code SNOOPY. Because there is no non-linear theory for inertial waves in vertically stratified boxes, we adopt an idealised purely local model which should capture the essential physics of the three-wave interactions. We expect the physics should be shared by vertically and fully global models. Our calculations are the first step in a larger program exploring the saturation of the VSI, with a focus on its wave-like character in global simulations, and the nonlinear wave dynamics that ensues.

Note that this paper complements the analysis in \citet{lp18}, which examines the stability and saturation of non-oscillatory VSI \emph{surface} modes; as we will show, the instabilities that attack the surface and the body modes are quite distinct. In particular, the predictions of \citet{lp18} are not applicable to the bulk of the disk, which are dominated by the travelling VSI wavetrains, as described above.

The paper is organized as follows. In \S\ref{sec:ge}, we present and justify our local model, while in \S\ref{sec:be} we give mathematical form to nonlinear inertial waves (VSI body modes) in this model. In \S\ref{sec:an}, we perturb these VSI waves and solve for the growth rates of parametric instability via the theory of three-wave resonance. In \S\ref{sec:nu}, we conduct numerical simulations with SNOOPY to examine the non-linear evolution of the inertial wave parametric instability. Finally, we discuss the main results in \S\ref{sec:disc} and summarise our findings in \S\ref{sec:cd}.

\section{Governing Equations}\label{sec:ge}

We adopt an idealised local model for a PPD, an incompressible shearing sheet \citep[see derivation in][]{lp17}, in order to highlight the main physics, i.e. the inertial wave character of the VSI, and its potential parametric instability. We thus examine the subsonic local dynamics of a small block of gas centred at $R=R_0$ and $Z=Z_0$ in a PPD, where $R$ and $Z$ are cylindrical radius and vertical height, and rotating with Keplerian angular velocity $\Omega = \Omega(R_0,Z_0)$. We erect a local Cartesian coordinate system with its origin located at the centre of the sheet, and in which $x$, $y$, and $z$ represent the radial, azimuthal, and vertical directions. The equations governing the motion of the gas in the sheet are:
\begin{align} \label{eq:ge1}
	&\pdv{\bu}{t} + \bu\cdot \nabla \bu = -\frac{1}{\rho}\nabla P - 2\Omega \be_z \times \bu + 3\Omega^2x \be_x +\nu\nabla^2\bu,   \\
	&\nabla \cdot \bu =0,  \label{eq:ge2}
\end{align} 
where $\bu$ is fluid velocity, $\rho$ is the constant density, $P$ is the pressure, and $\nu$ is the kinematic viscosity. We have omitted vertical shear, and as a consequence, the VSI cannot appear directly. It will, however, manifest indirectly as a background radially travelling inertial wave put in by hand (see \S\ref{sec:be}).

The above set of equations, admits steady equilibrium solution of linear orbital shear
\begin{equation} \label{eqm}
	\bu = \bu_0 = -\frac{3}{2}\Omega x \be_y,  \qquad P = P_0,
\end{equation} 
where $P_0$ is a constant. 

\section{Inertial Waves}\label{sec:be}

The incompressible local model is a particularly clean venue to explore inertial wave dynamics. As mentioned in the Introduction, these waves feature centrally in the VSI saturation in global simulations (where they are called body modes), but their essential properties can be captured in the shearing sheet. Moreover, the principle wave-wave couplings that redistribute the energy stored in the body modes and begin the turbulent cascade involve purely inertial waves. 
In this section, we outline the basic theory of such waves in our local model, and show how the global body modes excited by the VSI can be represented.

The steady state expressed in Eqs \eqref{eqm} is perturbed by axisymmetric disturbances $\bu_1$ and $P_1$, which obey
\begin{align} 
	\pdv{\bu_1}{t} & + \bu_1 \cdot \nabla \bu_1 = -\frac{1}{\rho}\nabla P_1 - 2\Omega \be_z \times \bu_1 + \frac{3}{2}\Omega u_{1x} \be_y,
\label{eq:5}	
\end{align} 
and 
\begin{equation} 
	\nabla \cdot \bu_1 =0.
\label{eq:6}	
\end{equation} 
We assume that the perturbations can be expressed as 
\begin{equation} 
	\bu_1, P_1 \propto \exp (\rmi \bk \cdot \bx - \rmi \omega t),
\end{equation} 
where $\bk = (\rmk_x, 0, \rmk_z)$, $\rmk_x$ and $\rmk_z$ are real wavenumbers, and $\omega$ is a real wave frequency. Due to  incompressibility $\nabla \cdot \bu_1 =\rmi \bk\cdot\bu_1 = 0$, the only non-linear term $\bu_1 \cdot \nabla \bu_1$ vanishes. As a consequence, linear solutions to the perturbation equations are also nonlinear solutions, and thus can take on arbitrary amplitudes.

In components, the equations \eqref{eq:5} and \eqref{eq:6} are
\begin{align} \label{lin1}
	-&\rmi \omega u_{x1} = 2\Omega u_{y1} - \rmi \rmk_x h_1,\\
	-&\rmi \omega u_{y1} = -\frac{1}{2}\Omega u_{x1},\\
	-&\rmi \omega u_{z1} = - \rmi \rmk_z h_1,\\
	&0 = \rmk_x u_x + \rmk_z u_z , \label{lin2}
\end{align} 
where $h_1=P_1/\rho$. This set of linearized equations produces a quadratic dispersion relation:
\begin{equation} 
	\omega^2 = \frac{1}{1+\rmk_x^2/\rmk_z^2} \Omega^2.
\label{eq:dr}	
\end{equation} 
The corresponding eigenmodes take the form
\begin{equation} 
\mqty(u_{x1}  \\ \\ u_{y1}  \\ \\ u_{z1}  \\ \\ h_1 ) =\epsilon \frac{\Omega}{|\rmk_z|}\mqty( 1  \\ \\ - (\Omega/2\omega) \rmi  \\ \\ -\rmk_x/\rmk_z \\ \\ -\omega\rmk_x/\rmk_z^2 ) \exp (\rmi \bk \cdot \bx - \rmi \omega t),
\label{eq:em}	
\end{equation} 
where we have chosen to scale the solutions by $\epsilon\Omega/|\rmk_z|$, with $\epsilon$ a tune-able dimensionless amplitude parameter. 

The inertial waves described above travel at an arbitrary direction in the $x$-$z$ plane, with the phase velocity directed along $\mathbf{k}$, and the group velocity directed perpendicular to $\mathbf{k}$. The body modes that appear in global simulations, however, travel in the radial ($x$) direction only, and manifest as standing waves in the $z$-direction: i.e., the vertical disk structure acts like a wave-guide \citep[e.g.,][]{lp93,kp95}. 
We can approximate this behaviour in our local model by summing two inertial waves counterpropagating in $z$. If $\mathbf{u}_+$ possesses the wavevector $\mathbf{k}_{+}=(k_x,0,k_z)$ and $\mathbf{u}_-$ the wavevector $\mathbf{k}_{-}=(k_x,0,-k_z)$, then our radially travelling `VSI wave' analogue is 
\begin{equation} 
	\mathbf{u}_\text{VSI}= \bu_+ +  \bu_-.
\label{eq:vsi}
\end{equation} 
Defining the phase variable
\begin{equation} 
\theta = \rmk_xx-\omega t, 
\label{eq:th}	
\end{equation} 
and retaining only the real part of $\bu_\text{VSI}$, we obtain
\begin{equation} 
	\Re [\mathbf{u}_\text{VSI} (x, z, t)] = \epsilon\frac{\Omega}{|\rmk_z|} \; \mqty( 2 \cos(\rmk_z z) \cos \theta \\ \\ \Omega/\omega\cdot  \cos(\rmk_z z) \sin \theta   \\ \\ 2 \rmk_x/\rmk_z \cdot \sin(\rmk_z z) \sin \theta).
\label{eq:bg}	
\end{equation} 
This wave solution will serve as part of our background (VSI body mode) in the following analytical and numerical work. 

Before continuing, note that Eq. (15) is not a nonlinear solution to our equations because of the non-zero quadratic terms $\mathbf{u}_-\cdot\nabla \mathbf{u}_+$ and $\mathbf{u}_+\cdot\nabla \mathbf{u}_-$, which introduce errors of order $\epsilon^2$. But, in fact, the parameter $\epsilon$ is expected to be small: in global simulations the body modes possess $\rmk_z\sim 1/H$, and thus $\epsilon$ can be interpreted as the Mach number ($u_1/c_s$) of the \emph{radial} velocity component. Typically simulations yield a Mach number for the vertical component of $\sim 0.1$ and for the radial of $\sim 0.01$ \citep[e.g.][]{nelson_etal13,sk14}. Thereby, the solution \eqref{eq:bg} is a reasonable approximation for our purposes.

\section{Stability of inertial waves}\label{sec:an}

In this section, we probe the dynamics of the radially travelling inertial wave described by Eq. \eqref{eq:bg}, which is serving as our local proxy for the VSI body modes that dominate global simulations. We begin by developing a theory for the three-wave coupling (or interchangeably, resonant interactions) between the VSI wave and two other smaller-scale inertial waves; the latter two can extract energy from the primary VSI via a parametric instability, and redistribute it to smaller scales (e.g. \citealp{craik86}). The theory works in the limit of a small amplitude of primary wave ($\epsilon \ll 1$), which also means the approximation \eqref{eq:bg} is valid. We also discuss how our analysis might form the basis of a weakly nonlinear dynamical system.

\subsection{Asymptotic analysis of parametric instability}%\label{sec:}

Our analysis here involves the interactions between three inertial waves. The primary is the background VSI body mode, which is assumed to be of small amplitude  (small $\epsilon$). The two secondary modes (labelled $A$ and $B$) are free modes and treated as linear perturbations upon the primary; if their frequencies and wavenumbers obey a set of resonance conditions, they can interact strongly with the primary wave, receive energy from it, and thus grow exponentially \citep[see, e.g.,][]{gammie00,latter16}. 

\subsubsection{The perturbation equations and asymptotic ordering}

We consider perturbations $\bv'=(v_x',v_y',v_z')$ and $P'$ upon a VSI inertial wave, so that $\bv= \mathbf{u}_0+\mathbf{u}_\text{VSI}+ \bv'$. In assuming that $\mathbf{u}_\text{VSI}$ is a solution to the governing equations, we have assumed that $\epsilon \ll 1$ in Eq.~\eqref{eq:bg}, and thus the VSI wave is of small amplitude. Errors introduced by this approximation come in at order $\epsilon^2$, which is too small to impact on the dynamics we examine.

The linearized equations governing the perturbations are
\begin{align} 
	&\pdv{\bv'}{t} + \bv' \cdot \nabla \bu_\text{VSI}  + \bu_\text{VSI} \cdot \nabla \bv'
	= -\frac{1}{\rho}\nabla h' + 2\Omega v_y'\be_x - \frac{1}{2}\Omega v_x'\be_y,  \label{eq:15}	\\
&\nabla \cdot \bv'=0, \label{eq:16}		
\end{align} 
where $h'=P'/\rho$.

The individual inertial wave oscillate on a `fast' timescale ($\lesssim\Omega$). In addition, the smallness of $\epsilon$ introduces a slow timescale into the problem, $\sim\epsilon\Omega$, upon which the wave-wave interactions and the parametric instability take place. To capture the slow evolution, we define a slow time variable $\tau=\epsilon t$, and write the perturbed quantities as
\begin{equation} 
	\bv' = \bv' (x,z,t,\tau), \qquad P' = P' (x,z,t,\tau).
\end{equation} 
By the chain rule, we have
\begin{equation} 
	\pdv{}{t} \to \pdv{}{t} + \epsilon\pdv{}{\tau}.
\label{eq:dt}
\end{equation} 
The following asymptotic expansions are then assumed for $\bv'$ and $h'$,
\begin{equation} 
\bv'=\bv'_1+\epsilon\bv'_2+..., \qquad h'=h'_1+\epsilon h'_2+.... %\qquad s=s_0+\epsilon s_1+....
\end{equation} 
These expressions are substituted into the linearized equations \eqref{eq:15}-\eqref{eq:16} and various orders in $\epsilon$ are collected. 

\subsubsection{Zeroth order and resonance condition}

At leading order $ O(\epsilon^0)$, the equations are
\begin{align} 
&\pdv{v'_{x1}}{t} -2\Omega v'_{y1} + \pdv{h'_1}{x} = 0, \\
&\pdv{v'_{y1}}{t} +\frac{1}{2}\Omega v'_{x1} =0, \\
&\pdv{v'_{z1}}{t} +\pdv{h'_1}{z}=0, \\
&\pdv{v'_{x1}}{x}+\pdv{v'_{z1}}{z}=0.
\end{align} 
These equations resemble Eqs \eqref{lin1}-\eqref{lin2}, and can be combined into $\mathcal{L}v'_{x1}=0$, where the `inertial wave' differential operator is (e.g. \citealp{tl21})
\begin{equation} 
\mathcal{L}=\pdv{^2}{t^2}\bigg( \pdv{^2}{x^2}+\pdv{^2}{z^2}\bigg) +\Omega^2\pdv{^2}{z^2}.
\end{equation} 
This differential equation is straightforward to solve because it is linear and has constant coefficients. Assuming $v'_{x1}\propto \exp(\rmi \bk \cdot\bx-\rmi \omega t)$ gives
\begin{equation} 
\omega^2=\frac{\rmk_{z}^2}{\rmk^2}\Omega^2,
\label{eq:s1}
\end{equation} 
Clearly, the leading order analysis results simply in inertial wave oscillations, as described in Section 3. One can also write down the expressions for $v'_{x1},v'_{y1}, v'_{z1}, h'_1$, as in Equation \eqref{eq:em}. 

The general solution at this order is a linear superposition of all the inertial waves of different $\rmk_x$ and $\rmk_z$ described above. We select of this infinite set just two waves that can come into resonance with the primary VSI wave. We denote their wavevectors by $\bk_A$ and $\bk_B$, their oscillation frequencies by $\omega_A$ and $\omega_ B$, and their velocity eigenfunctions by $\mathbf{v}_A$ and $\mathbf{v}_B$. Thus we set
\begin{equation} \label{defn}
v'_{x1} = A(\tau)E_A + B(\tau)E_B 
\end{equation}
where 
\begin{align} 
E_A=\exp(\rmi \bk_A\cdot\bx-\rmi \omega_A t), \quad E_B=\exp(\rmi \bk_B\cdot\bx-\rmi \omega_B t),
\end{align} 
$A$ and $B$ are complex amplitude functions of the slow variable. The oscillation frequencies are determined from the dispersion relation \eqref{eq:s1}, and can be of either sign. 

In order for the $A$ and $B$ waves to couple with the underlying VSI wave, their wavevectors and frequencies must satisfy resonance conditions:
\begin{equation} 
\bk_\text{VSI} \pm \bk_A \pm \bk_B=\mathbf{0}, \qquad \omega_\text{VSI} \pm \omega_A \pm \omega_B= 0,
\label{eq:rc}
\end{equation}
where $\bk_\text{VSI}$ and $\omega_\text{VSI}$ are the wavevector and frequency of the background VSI. In the following we take the signs to be `$+$' and `$-$' in front of $\bk_A$ and $\bk_B$ in Eq.~\eqref{eq:rc}

Before moving to the next order, recall that the background VSI wave is a superposition of two waves counterpropagating in the vertical direction, proportional to either $E_+$ or $E_-$, where 
\begin{align} 
E_\pm=\exp(-\rmi\omega_\text{VSI} t+\mathbf{k}_\pm\cdot\mathbf{x}).  
\end{align} 
One may then use the resonance conditions to derive the useful identities:
\begin{equation} \label{identities}
E_A E_+=E_B, \qquad E_B E_+^*=E_A.
\end{equation}
Note that other quadratic products of $E_A$ and $E_B$ with $E_+$, $E_-$ and their complex conjugates cannot equal $E_A$, or $E_B$, and hence will not play a role in what follows.

\subsubsection{First order and growth rate formula}

At the next order $O(\epsilon)$ we derive the following equations, 
\begin{align} 
&\pdv{v'_{x2}}{t} - 2\Omega v'_{y2} + \pdv{h'_2}{x} = F_x, \\
&\pdv{v'_{y2}}{t} + \frac{1}{2}\Omega v'_{x2} = F_y,\\
&\pdv{v'_{z2}}{t} + \pdv{h'_2}{z} = F_z,\\
&\frac{\p v'_{x2}}{\p x}+\frac{\p v'_{z2}}{\p z}=0,
\end{align}
where 
\begin{equation}
\bF = -\pdv{\bv'_1}{\tau}-\bv'_1\cdot\nabla\bu_\text{VSI}-\bu_\text{VSI}\cdot\nabla\bv'_1.
\end{equation} 
The above set of equations can be reduced to the convenient inhomogeneous differential equation $\mathcal{L}v'_{x2}=G$, where the `forcing term' is
\begin{align} 
G = \frac{\p^3F_x}{\p t \p z^2} + 2\Omega\frac{\p^2F_y}{\p z^2}
-\frac{\p^3F_z}{\p t\p x\p z}.
\end{align}
In order for the asymptotic expansion to remain valid (i.e. $|\mathbf{v}'_1|\sim 1$), a solvability condition is required, namely that $G$ has no component proportional to a solution of the homogeneous problem. In our case, this means that $G$ cannot be proportional to either $E_A$ or $E_B$. Using \eqref{identities} we can manipulate $G$ into
\begin{align}\label{GG}
G = g_A E_A + g_B E_B + ... \ ,
\end{align} 
where the ellipsis contains other terms that are not eigenfunctions of $\mathcal{L}$. The coefficients $g_A$ and $g_B$ involve $A$ and $B$ and their $\tau$ derivatives. These coefficients must be set to 0 to satisfy solvability, which then produces two ODEs:
\begin{equation}
\frac{dA}{d\tau}-\text{i}a B=0, \qquad
\frac{dB}{d\tau}+\text{i}b A =0,
\end{equation}
where the constants $a$ and $b$ are given by
\begin{align*}
    a&= \frac{1}{2}\omega_A (\mathbf{k}_+\cdot \mathbf{v}_B)\left[\omega_A\frac{(\mathbf{k}_+\cdot \mathbf{k}_A)}{\rmk_{Az}}-\frac{1}{\omega_\text{VSI}}\right] \\ 
    &\hskip1.75cm -
    \omega_A (\mathbf{k}_B\cdot \mathbf{u}_+^*)\left[\omega_A\frac{(\mathbf{k}_A\cdot \mathbf{k}_B)}{\rmk_{Az} \rmk_{Bz}}+\frac{1}{\omega_B}\right], \\
    b&= \frac{1}{2}\omega_B (\mathbf{k}_+\cdot \mathbf{v}_A)\left[\omega_B\frac{(\mathbf{k}_+\cdot \mathbf{k}_B)}{\rmk_{Bz}}+\frac{1}{\omega_\text{VSI}}\right] \\
    &\hskip1.75cm +\omega_B (\mathbf{k}_A\cdot \mathbf{u}_+)\left[\omega_B\frac{(\mathbf{k}_A\cdot \mathbf{k}_B)}{\rmk_{Az}\rmk_{Bz}}+\frac{1}{\omega_A}\right],
\end{align*}
On assuming that $A,\, B \propto \exp(s\tau)$, where $s$ is the growth rate of the parametric instability with respect to the slow timescale, we obtain 
$s = \sqrt{ab}$. Thus the actual parametric growth rate in dimensional units is $\epsilon \sqrt{ab}\,\Omega$.

A more convenient expression of the growth rate can be obtained in the small wavelength limit of the secondary modes, which is also the regime of fastest growth. If we assume both their radial and vertical wavenumbers are much larger than the underlying VSI wave's, then the resonance condition tells us that $\mathbf{k}_A\approx \mathbf{k}_B$, and furthermore that
\begin{align}
\frac{\rmk_{Ax}}{\rmk_{Az}}\approx \sqrt{3+4K^2}, \qquad \omega_A=-\omega_B=-\frac{1}{2}\omega_\text{VSI},
\end{align}
where we have introduced $K=\rmk_{\text{VSI},x}/\rmk_{\text{VSI},z}$ for notational convenience.
The growth rate formula in this limit simplifies to 
\begin{align}\label{qmax}
s^2 = \frac{27 + 54 K^2 + 27 K^4 + 8 K^6 + 4K^3(3+ K^2)\sqrt{3+4K^2}}{64 (1 + K^2)^2}.
\end{align}
In the case representative of VSI body modes, $K$ is large, and we have to leading order
$s\sim \frac{1}{2}K$; thus the parametric growth rate in physical units is $\sim \frac{1}{2}K\epsilon \Omega $, though strictly $K$ must remain significantly smaller than $1/\epsilon$ for the asymptotic analysis to remain valid. Nonetheless, it is clear that the parasitic modes that attack more elongated VSI waves grow more vigorously.   

\subsubsection{Illustrative examples}\label{sec:414}

In order to illustrate the previous calculations we select two different underlying VSI waves with wavevector orientations: $K=\rmk_{\text{VSI},x}/\rmk_{\text{VSI},z}=1$ and $K=10$. The latter is a configuration we expect to be more likely favoured by the VSI.

The resonance condition is satisfied by several wave triads. If we restrict the vertical wavenumbers of the secondary inertial waves to be integer multiples of $\rmk_{\text{VSI},z}$, then resonance occurs for an infinite set of pairs of $\mathbf{k}_A$ and $\mathbf{k}_B$. The first few triads in both $K$ are displayed in Table \ref{table:1}, alongside the associated parametric growth rate. For comparison, the growth rate in the limit of large wavenumber is 
$s_\text{max}=\frac{1}{16}\sqrt{116+\sqrt{7}}\approx 0.786438$ for $K=1$, and $s_\text{max}\approx 5.03430$ for $K=10$, as predicted from Eq.~\eqref{qmax}. In fact, every triad grows at a similar rate, and the formula \eqref{qmax} serves as a good general approximation. 

\begin{table}
\begin{tabular}{ | l | l | l | l |}
 \hline
  $\bk_\text{VSI}$ & $\mathbf{k}_A$   &  $\mathbf{k}_B$  & $s$  \\
 \hline
\hline
$(1,1)$ & $(3.32893,1)$   
  & $(4.32893,2)$ & 0.770217 \\
 \hline
$(1,1)$ & $(6.03214,2)$   
 & $(7.03214,3)$ & 0.780814  \\
\hline
$(1,1)$ & $(8.70169,3)$ 
 & $(9.70169,4)$ &  0.783597 \\
\hline
$(1,1)$ & $(11.3605,4)$ & $(12.3605,5)$  & 0.784726 \\
\hline
$(1,1)$ & $(14.0146,5)$ & $(15.0146,6)$  & 0.785294 \\
\hline
$(1,1)$ & $(27.261,10)$
  & $(28.261,11)$  & 0.786125 \\
 \hline
$(1,1)$ & $(133.106,50)$
 & $(134.106,51)$ & 0.786424
  \\
\hline
\hline
$(10,1)$ & $(24.248,1)$  & $(34.248,2)$ & $4.95919$
  \\
  \hline
  $(10,1)$ & $(44.6783,2)$  & $(54.6783,3)$  & $5.00838$
  \\
  \hline
$(10,1)$ & $(64.9003,3)$ & $(74.9003,4)$ & $5.02122$
  \\
  \hline
$(10,1)$ & $(85.0561,4)$ & $(95.0561,5)$ & $5.02642$ 
  \\
  \hline
  $(10,1)$ & $(105.182,5)$ & $(115.182,6)$  & $5.02904$ 
  \\
  \hline
$(10,1)$ & $(205.666,10)$ & $(215.666,11)$ & $5.03286$
  \\
  \hline
$(10,1)$ & $(1008.76,50)$  & $(1018.76,11)$  & $5.03423$
  \\
  \hline
\end{tabular}
\caption{Wavevectors of selected resonant triads for two classes of VSI wave, $K=1$ and $K=10$. Note that $\bk=({\rmk_x,\rmk_z})$ is shown, and the $y$-component of all the wavevectors is 0, thus omitted. The associated parametric growth rate $s$ is also shown. }
\label{table:1}
\end{table}

%\subsubsection{`Double resonance'}

%We have restricted the possible resonances that the underlying VSI wave can initiate. In actual fact, growing secondary waves can participate in two resonances at the same time through other combinations of the signs in the condition \eqref{eq:rc}. For instance, mode $B$ can fall into resonance with the modes $(\mathbf{k}_A,\omega_A)$ and $(\mathbf{k}_A,-\omega_A)$, with each producing a growth rate of the same magnitude. The net effect of these overlapping or `double' resonances is that the growth rates computed earlier are doubled.  We emphasise this is not a case of a four-wave resonance (see \citealp{reun2020} for further discussion). In the larger context, this effect is not especially important, but it should be accounted for when comparing the theory with numerical simulations. 

\subsection{Nonlinear development and saturation}
\label{nlsat}

The parametric instability removes energy from the primary wave and redistributes it among the two secondary modes. In the absence of energy input via the VSI, and loss via viscosity, the total energy will be conserved, and the system will relax to a quasi-steady state (e.g. \citealp{craik86}). But it is possible to account for the input and loss, and thus to describe the VSI saturation behaviour in an ad hoc way, which is our goal in this subsection. 

\subsubsection{A simple `parasitic' theory}

One approach is to assume that at saturation, the energy input into the primary wave by the VSI must be balanced by its removal by the parametric instability. This can be modelled crudely by equating the VSI and parametric growth rates.  The local maximum VSI growth rate is $\approx |q|\Omega$, where $q=R_0(\partial \ln\Omega/\partial Z)_0\sim H/R\ll 1$, and occurs for $K\sim 1/q\gg 1$. The parametric growth rate, on the other hand, is $\approx \frac{1}{2}K\epsilon\Omega$, when $K$ is large. If these rates are of the same order, we can solve for the amplitude of the underlying VSI wave: $\epsilon\sim q^2$. Furthermore, taking the VSI's $k_z\sim 1/H $, then its radial Mach number is $\sim q^2\sim (H/R)^2$, while its vertical and azimuthal Mach numbers are $\sim q\sim H/R$ (from inspection of its eigenmode structure, cf. Eq.~\ref{eq:em}). It follows that the alpha parameter is $\sim q^3 \sim (H/R)^3$, in agreement with a scaling derived in \citet{bl15} using a different argument. Note that the vertical flux of angular momentum, associated with the $y$-$z$ component of the Reynolds stress, will be a factor $R/H$ larger than the radial flux. 

Note that, because inertial waves are relatively insensitive to the disk thermodynamics, the cooling physics of the gas does not enter directly in these saturation estimates. The cooling time can appear implicitly, however, via the VSI growth rate. In the above calculations, this was taken to be simply $|q|\Omega$, but if its dependence on cooling was included then that dependence would be passed on, in some fashion, to the estimated saturation amplitude.  

Recently, a parameter survey conducted by \citet{manger_etal20} showed that the saturated amplitudes of the total rms velocities do obey $u_\text{rms}\sim c_s (H/R)$. On the other hand, the Reynolds stress scales as $\sim c_s^2(H/R)^{2.6\pm 0.3}$, which is marginally consistent with our estimate. In fact our steeper scaling leads to a better magnitude match with the numerical data, as the multiplicative constant remains of order one in most cases. However, as will be discussed in more detail in Section \ref{sec:ds1}, the resolution of these simulations is very coarse and it is unlikely that the parametric instability is working in these simulations. The reasonable agreement we find is then probably fortuitous. 
%A final point concerns the dependence of the saturation

\subsubsection{Weakly nonlinear theory}

One can do a little better if it is assumed that the amplitudes of the coupled waves remain small. It is then possible to derive a dynamical system describing the evolution of the wave amplitudes of the three modes. While the viscous and nonlinear terms arise naturally from our local modes, we must put in by hand the energy input from the VSI. \footnote{In a suitable vertically stratified model \citep[e.g.][]{nelson_etal13} it should be possible to derive the nonlinear system \eqref{nl1}-\eqref{nl3} completely rigorously. This forms the basis of future work.} The system is 
\begin{align}\label{nl1}
\quad \frac{dA}{d\tau} = -\mu_A A+ \text{i}a B  U^*, \qquad &\frac{dB}{d\tau} = -\mu_B B- \text{i}b A  U, \\ \frac{dU}{d\tau}= \sigma U + &\text{i}c A^*B, \label{nl3}
\end{align}
where we denote the underlying VSI wave's complex amplitude to be $U$, $\mu_A$ and $\mu_B$ represent viscous damping, $\sigma$ is the VSI growth rate, and $c$ is a real constant that can be rewritten in terms of $a$ and $b$ using conservation of energy. The behaviour of this system is well understood (e.g. Vyshkind and Rabinovic 1976, \citealp{Wersinger+80}), and permits various bounded oscillatory states, including those where the primary VSI dominates, but its growth is halted via energy redistribution to the damped secondary modes. These states are organised around a fixed point in the phase space, which consequently sets the amplitude of the saturation. This fixed point and the expected saturation amplitudes can be calculated, and they are in rough agreement with the argument based on equating growth rates above (see Appendix A in \citealp{tl21}). 

\subsubsection{Wave turbulence and zonal flows}

Finally, it should be clear from Table \ref{table:1} that multiple triads could be working concurrently, each producing a distinct cascade of mode couplings. Thus, the simple picture described by Eqs \eqref{nl1}-\eqref{nl3} is only one component in a potentially complicated dynamics and, as a consequence, the saturation estimates posited above might have to come under significant revision. It is worth remarking that via resonant interaction of this type, the primary mode can directly send energy to lengths many orders of magnitude smaller than itself, vaulting all the intervening scales. 

An alternative statistical description of the ensuing chaotic flow is provided by the Kuznetsov-Zhakarov weak/wave turbulence theory (e.g. \citealp{galtier03,ns11}), which may have some application in the VSI context. A further interesting phenomenon is the generation of large-scale coherent structure via nonlinear mode couplings, namely zonal flows and associated vortices (e.g. \citealp{SW99,K99,reun2020}). We leave to future work an application of these ideas to VSI turbulence.

\section{Numerical simulations}\label{sec:nu}

We conduct numerical simulations to confirm and extend the theory of three-wave resonant interactions developed in the preceding section. 

\subsection{Method and Simulation Setup}

Numerical simulations are performed via the spectral code SNOOPY\footnote{\url{https://ipag.osug.fr/~lesurg/snoopy.html}} in the incompressible, axisymmetric shearing sheet, solving Equations \eqref{eq:ge1}-\eqref{eq:ge2}. Working in Cartesian coordinates ($x, z$), the simulation domain spans $[-L/2,L/2]$ in both dimensions, where $L$ is the box size. We adopt periodic boundary conditions. The resolution is set at $64\times 64$ for $x\times z$, which is fine enough to resolve the parametric instability in the present work. The rotational profile is Keplerian, and no vertical shear is implemented. Code units are taken so that $\Omega=1$ and $L=1$. The strength of viscous diffusion is parametrized by the Reynolds number Re$=L^2\Omega/\nu$, and is set to $\mathrm{Re}=10^6$.

 Since the simulations are performed in a local shearing box model, whereas the VSI body modes only arise in a vertically global model, we set up an initial value problem with a proxy for the background VSI wave (Equation \eqref{eq:bg}). The wavenumbers of the background VSI wave are set to be $k_{\rm VSI, x}=2\pi, k_{\rm VSI, z}=2\pi$, such that the simulation box contains exactly one wavelength in each direction. Though in global simulations $k_{\rm VSI, x}/k_{\rm VSI, z}$ is large, we find little difference in our results when varying this ratio. Informed by our discussions in Sections 3 and 4.2, we adopt a primary wave amplitude of $\epsilon=5\times10^{-3}$. 
In addition, to initiate the parametric instability, random noise is superposed on the background VSI wave, with an amplitude of 1$\%$ of the background wave amplitude $\epsilon$. Lastly, we run the simulation up to $400P$, where $P=2\pi/\Omega$ is the orbital period.
 
\subsection{Simulation Results}\label{sec:sr}

\subsubsection{Radial propagation of pure VSI wave}\label{sec:521}

We first examine the behaviour of the background inertial wave \eqref{eq:bg}. This is done with white noise omitted. The wave propagates through the box for $400P$ and retains its coherence throughout. The radial phase velocity of the background inertial wave is $v_{\rm phase, x}=\omega/k_{\text{VSI},x} =\sqrt{0.5}/(2\pi)$. We calculate the phase velocity numerically from the simulation, compare with the theory, and find that by $400P$ the error is $\sim0.5\%$, which is acceptably small, and should not pollute the main results below. 

\begin{figure*}
\centering
\includegraphics[width=1\textwidth]{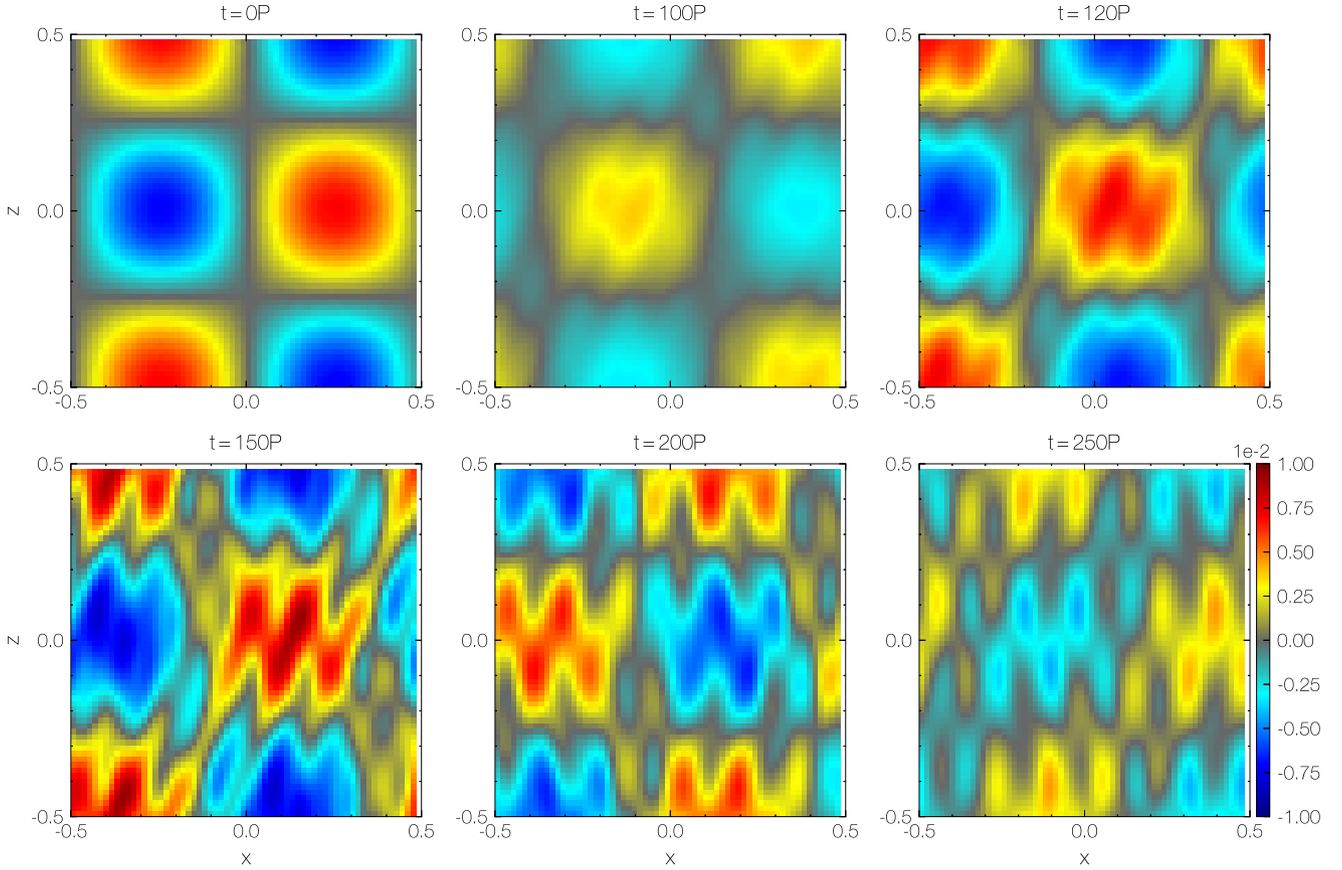}
\caption{Snapshots of azimuthal velocity $v_y$ at $t=0,100,120,150,200,250P$.}
\label{fig:ss}
\end{figure*}

\begin{figure*}
\centering
\includegraphics[width=1\textwidth]{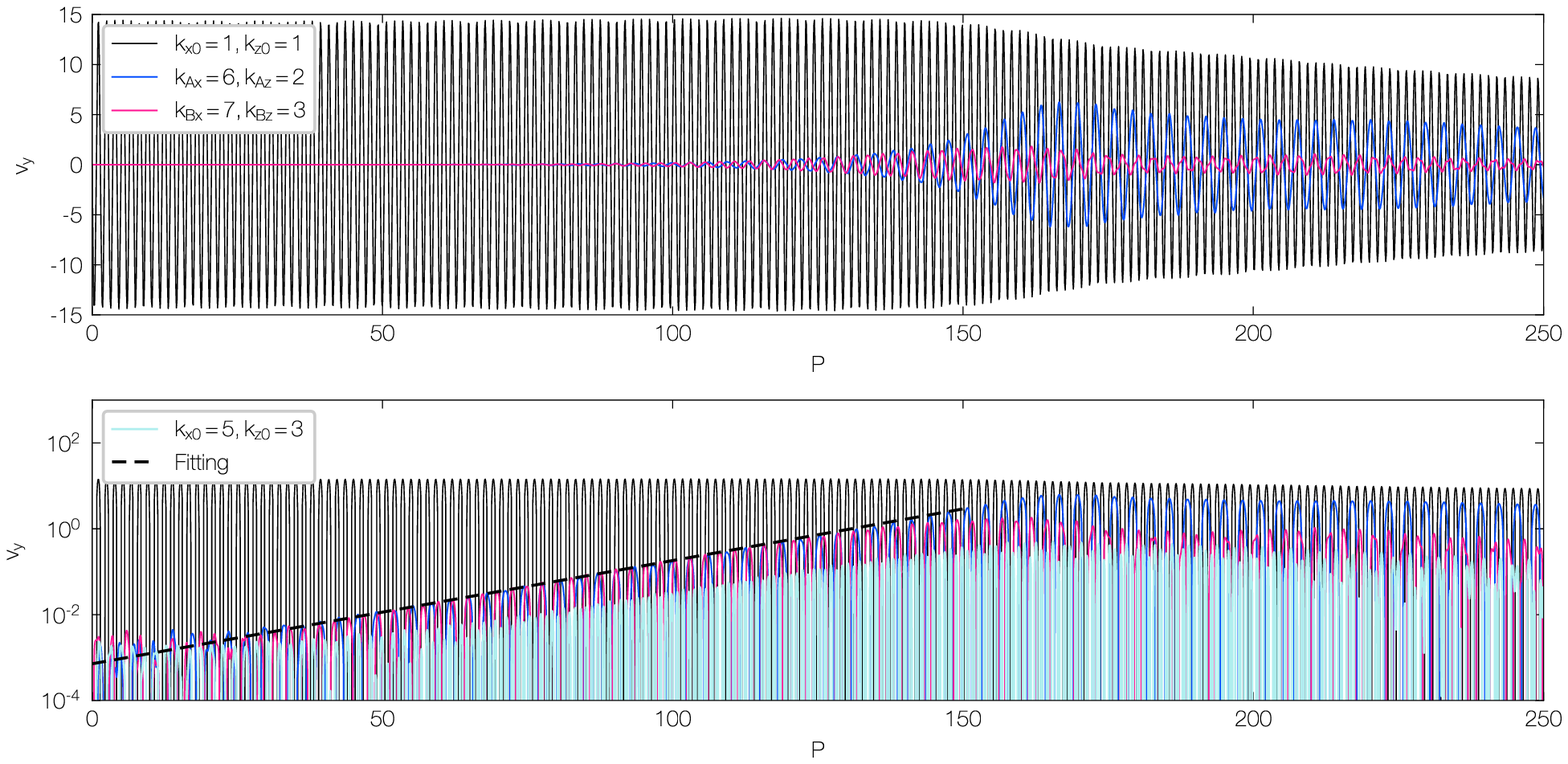}
\caption{Azimuthal velocity $v_y$ as a function of time for different wave modes. Top panel: the black curve denotes primary VSI wave with $(k_{\text{VSI},x},k_{\text{VSI},z})=(1,1)$. The blue and red curves denote secondary inertial waves with $(k_{Ax},k_{Az})=(6,2)$ and $(k_{Bx},k_{Bz})=(7,3)$. Bottom panel: the light blue curve denotes another mode $(k_x,k_z)=(5,3)$ that participates in the resonance. The black dashed line denotes a fitting of the linear growth rate of secondary mode A (blue curve).}
\label{fig:modes}
\end{figure*}

\subsubsection{Linear growth of parametric instability}

We then introduce white noise on top of the background VSI wave. Figure \ref{fig:ss} shows snapshots of azimuthal velocity $v_y$ at $t=0,100,120,150,200$, and $250P$. At $t=0P$, we can observe the primary inertial wave, and the amplitude of the white noise is too small to observe. At $t=100P$ and $200P$, the parametric instability begins to develop, though the structure of the primary wave still dominates. In the following snapshots, the structure of the parametric instability become more readily discernible. By eye, one of the secondary wave modes likely has a radial wavenumber of $6$ and a vertical wavenumber of 2, seen clearest at $t=250P$.

To quantitatively determine which secondary modes are amplified in the simulation, in Figure \ref{fig:modes} we decompose the azimuthal velocity $v_y$ into Fourier components, and examine Fourier amplitudes as a function of time. Note that Fourier amplitudes are unnormalized, and thus only meaningful in relative terms. The top panel presents the real amplitudes of three dominant modes. These modes are represented by wavevectors, normalised by $k_{\text{VSI},z}=2\pi$. The black curve denotes the primary wave with $k_{\rm VSI, x},k_{\rm VSI z}=(1,1)$, the blue curve represents the mode with $k_{Ax},k_{Az}=(6,2)$ and red with $k_{Bx},k_{Bz}=(7,3)$.  We see that all three modes oscillate on a fast timescale corresponding to their inertial wave period $\gtrsim P$, while varying slowly on a longer timescale $\sim 150P\sim P/\epsilon$, as predicted in Section 4.1. The two secondary modes grow exponentially on this longer timescale, which is the timescale for the linear parametric instability. By $t=150P$, the secondary modes have reached the same order of magnitude of the primary wave amplitude.  

To verify whether these amplified secondary wave modes are the ones expected from the theory, we turn to Table \ref{table:1}, which lists various resonant triads that might operate in the case simulated, i.e. $(k_{\rm VSI, x},k_{\rm VSI, z})=(1,1)$. We find the pair $(6.03214,2)$, $(7.03214,3)$, which correspond to the growing modes observed. The reason why other triads are not seen is because of the periodic boundary conditions adopted. If the resonant modes possess wavenumbers that are insufficiently close to integer multiples of $k_{\text{VSI},z}=2\pi/L$, then they cannot fit into the box. The next nearest resonant modes that could grow are the pair $(14.0146,5)$, $(15.0146,6)$. Their growth rate is marginally greater than the longer pair, but their short lengthscale means they are easily diffused away. Thus, we conclude that the theory and simulation are in overall good agreement in terms of the resonance condition.
 
The bottom panel of Figure \ref{fig:modes} is idential to the top panel, but adopting a logarithmic scale in y-axis. We fit the growth rate of the secondary wave (mode A) with a black dashed line, which indicates a clear exponential growth. The numerical growth rate can then be calculated, yielding $s=1.795$. This must be compared with the theoretical growth rate $s=2\times 0.7808=1.562$; the factor of 2 comes about because each mode in the resonance partakes in a `double resonance' \citep[see discussion in][]{reun2020}. The numerical value is some $10\%$ larger than the theoretical, a discrepancy that motivates us to check if we are in the regime of sufficiently small $\epsilon$; if $\epsilon$ is large, then other wave modes may potentially couple together and enhance the growth rate \citep[see, e.g.,][]{mied76, drazin77}. Indeed, in the simulation we find that $(3,3), (5,1), (5,3), (6,4)$ and $(7,1)$ modes grow at exactly the same rate as the dominant $(6,2)$ and $(7,3)$ modes. Among them, we plot (5,3) in the bottom panel of Figure \ref{fig:modes}. Thus, the theory of Section 4.1 is not fully valid here. Unfortunately, if we reduce $\epsilon$ to smaller values, numerical and physical viscosity intervene, and we witness no parametric growth at all, making a more exact comparison between theory and simulation impossible. Nonetheless, overall the agreement is sufficient to demonstrate our main idea.

\subsubsection{Nonlinear development of parametric instability}\label{sec:523}

It is clearly seen that the non-linear effects start from $\sim 150P$ in the bottom panel of Figure \ref{fig:modes}. From this point, the amplitude of the primary wave declines over time, as it conveys its energy to the secondary waves. Subsequently, the energy of the system as a whole decreases slowly due to viscosity. The secondary modes are responsible for the majority of these losses, as they possess shorter length-scales. In a realistic PPD, however, there is an energy source from the vertical shear to sustain the VSI body modes, unlike our simulations whose energy is entirely contained in the initial conditions. We expect potentially a steady balance between the VSI injection and viscous losses in real disks (see Section 4.2.2).

\section{Discussion}\label{sec:disc}

\subsection{Global simulations and resolution requirements}\label{sec:ds1}

In this subsection, we apply the theory of parametric instability (Section \ref{sec:an}) to current VSI global simulations, focusing on the resolution needed to resolve the parametric instability. Our calculations (Table \ref{table:1}) suggest that the lowest-order resonant modes possess a radial wavelength about $1/3$ that of the VSI body mode. Whether a global simulation has the capacity to resolve the secondary wave modes depends on 1) the natural radial wavelength of the VSI body mode $\lambda_r$, and 2) how many grid points there are per $\lambda_r$. For the discussion below, we assume that about $10$ grid points are necessary to adequately resolve one wavelength.
 
 As regards the first point, there is still no secure theory on which VSI body modes wavelengths are selected in global simulations. Assuming the dominant body modes are the fastest growing ones, linear theory suggests $\lambda_r \sim q H \sim (H/R)H$ \citep{lp18}. If this is true, it would require a punishing resolution, $\sim100$ grid cells per $H$ in radius, for global simulations to just resolve the primary VSI wave. To further resolve the longest wave modes of the parametric instability, it requires at least $\sim300$ grid cells per $H$ in radius, which no simulation has yet reached. An even more pessimistic viewpoint might assume that the parametric instability destroys body modes, which would mean simulations will always settle on an (artificially stable) body mode with radial wavelength near the grid length. The ensuing saturated state would then be qualitatively incorrect. 
 
 Thankfully, there is evidence that the scaling $\lambda_r \sim (H/R)H$ assumed above may be too strict. In an interesting observation, \citet{sk14} show that the radial wavelength increases with radius more steeply than this scaling (see their Figure 8). Thus, as the body modes propagate radially away (outwards) from where they are launched, they lengthen considerably and become better resolved. Furthermore, recent very high resolution simulations by \citet{flores_etal20} reveal dominant body modes with a radial wavelength resolved by $\sim 80$ grid points, which suggests both they and their parametric instability are resolved.
In particular, radiative diffusion and non-ideal MHD effects seems to increase the dominant radial wavelength \citep{flores_etal20,cb20}. In summary, high resolutions appear to be necessary but $\sim300$ cells per H might be an overestimate.
 
 Future work should concentrate on determining the natural wavelength of the preferred VSI body modes, and to determine if it is possible to obtain convergence with numerical resolution, and whether the longest parametric instability can be described.
 Even then it is likely that only limited number of resonant triads will appear: as shown in Table \ref{table:1}, there is a large number of modes participating in resonant interactions. Their omission in the small-scale dynamics may impact especially on how faithfully dust settling, collisions, and concentration are reproduced. 

\subsection{Feedback of the VSI on the vertical shear}

In Section 4.2, we obtained estimates for the VSI's saturated flow speeds in the case of a fixed vertical shear, i.e. constant $q=R_0(\partial\ln\Omega/\partial Z)_0$. Specifically, $u_r\sim q^2 c_s$ and $u_\phi,\,u_z\sim q c_s$, where $c_s$ is the sound speed. However, we expect the VSI will attempt to erase the destabilising conditions from which it arose, and thus act back upon $q$ so that it is reduced as close to zero as possible. This may be effected by altering the vertical thermal structure of the disk, through viscous heating and the vertical transport of heat, or by the vertical redistribution of angular momentum. Both routes will be resisted by the thermal driving of the disk from the protostellar radiation field, and the consequent poloidal circulations that establish thermal wind balance. In this section, we assess how successful the VSI is in acting back on the disk's unstable equilibrium, and conclude that it is unlikely to do so significantly.     

\subsubsection{Thermal structure}

Generally, internal hydrodynamic activity struggles to reshape the vertical thermodynamic structure of a protoplanetary disk. Turbulent heating beyond a critical radius $\approx 2$ AU is a  subdominant effect in the main hydrodynamic balances, which are instead controlled by irradiation from the central star (e.g. \citealp{d98}). Meanwhile, the vertical turbulent transport of heat by convection, a popular idea in earlier accretion disk research, is relatively limited \citep{cassen93,rafikov07} and we expect thermal VSI transport to be weaker than convection.

Perhaps more fundamentally, in order for the VSI to reshape the disk's vertical structure, the thermal relaxation/cooling time of the gas $\tau_\text{relax}$ must be sufficiently long, certainly no shorter than the characteristic VSI timescale $(q\Omega)^{-1}$. But if $\tau_\text{relax}$ were as long as that, the VSI would fail to grow in the first place: to circumvent the stabilising influence of stratification, the VSI requires $\tau_\text{relax}\ll q (\Omega/N^2)\sim q\Omega^{-1}$ \citep[see linear estimates in:][]{ly15,lp18}. 
All in all, it is relatively clear that the VSI will be unable to feedback on to the thermal state which instigated the vertical shear. 

\subsubsection{Angular momentum redistribution}

The VSI might fare better in reshaping the vertical shear itself. The $z\phi$ component of the specific Reynolds stress is $\sim c_s^2 q^2$. In the absence of baroclinic forcing, it will smear out the deviation from cylindrical rotation $R_0\Delta \Omega\approx R_0\Omega q (H/R) $ on the VSI's characteristic timescale $\tau_\text{VSI}\sim (q\Omega)^{-1}$, as can be verified by balancing terms in the $\phi$ component of the momentum equation, and recognising $q\sim H/R$. 

 Countering this process is a poloidal circulation driven by the disk's baroclinicity, which will transport angular momentum vertically so as to restore the shear.
 %\footnote{As explained in \citet{ks86}, this flow has little impact on the thermal structure of the disk, and can be thought of the `converse' of the Eddington-Sweet circulation in stellar interiors.}.
 If $\varpi$ is the toroidal component of this flow's vorticity, and $\tau_\text{bc}$ is the characteristic timescale of the momentum redistribution, then we have $$\frac{\varpi}{\tau_\text{bc}}\sim -\frac{(\nabla\rho\times\nabla P)_\phi}{\rho^2}\sim \frac{c_s^2}{R H},$$  from the $\phi$ component of the vorticity equation. Furthermore, setting $\varpi\sim  1/\tau_\text{bc}$, we have $\tau_\text{bc}\sim q^{-1/2}\Omega^{-1}$, which we take to be the natural timescale that the vertical shear is established. 
 
 The VSI timescale is longer than the baroclinic timescale by a factor $q^{-1/2}\sim (R/H)^{1/2}\sim 3-4$. We hence expect that, at best, the VSI erodes the vertical shear by some small but non-negligible factor. To a first approximation, it may be acceptable to take $q$ as a fixed quantity that remains mostly unchanged by the VSI. Our estimates for saturation amplitudes should then hold, though the above argument is certainly rough and needs to be tested in numerical simulations.

\section{Conclusion}\label{sec:cd}

In this paper, we study the saturation process of the VSI mediated by a parametric instability. The dominant VSI modes in a global disk model are body modes, which are essentially radially travelling inertial waves. These waves come into triadic resonances with pairs of smaller scale inertial waves, and the ensuing parametric instability transfers energy from the VSI body modes into the secondary waves, where it could initiate an inertial-wave turbulent cascade. Subsequently, a steady state can form, with energy input from the VSI body modes balanced by the energy losses on the viscous scales. 

We start with establishing a linear theory for the parametric instability. We employ an idealised local incompressible model, which cleanly isolates the three-wave coupling and produces analytic formulae for the parametric growth rate. 
Next, we conduct numerical simulations with the SNOOPY code to verify the theory and find a good agreement between the two.
Our results predict a far more incoherent flow pattern than what current global simulations suggest, which can have meaningful implications for dust dynamics. 
Furthermore, we indicate that the resolutions in current global simulations might be insufficient to properly resolve the parametric instability, which may require $\sim 300$ grid cells per H in radius.  

Future work might take several directions. First, the analytic calculations in the present paper could be extended to a vertically stratified box, which would include energy input into the body modes self-consistently. Currently there is no theory for inertial waves in disks that achieve non-linear amplitudes, but this is the form that the VSI takes in global simulations. It is hence a priority to develop such a theory.
In addition, high resolution global simulations need to be analysed specifically to recover evidence of parametric instability, and the resonant waves predicted by the theory. And, if numerically feasible, higher resolution runs are needed to make sure that when more resonant triads appear in the simulations, the saturation properties do not qualitatively change. Finally, we must determine the properties of the small-scale inertial wave turbulence generated by the parametric instability, possibly via local models. Initially this turbulence will be axisymmetric, but there is the likelihood that nonlinear wave interactions form larger coherent structures, such as zonal flows and vortices, as witnessed in other similar contexts \citep{bl13,tl21}, and which certainly appear in global VSI simulations \citep[e.g.][]{richard_etal16}.

\section*{Acknowledgements}

We thank the anonymous referee for a very prompt report, and Adrian Barker for a set of useful comments. We thank Lorenzo Perrone for helpful discussions on the SNOOPY code. CC and HNL acknowledge funding from STFC grant ST/T00049X/1. 

\section*{Data Availability}

The data underlying this article will be shared on reasonable request to the corresponding author.

%%%%%%%%%%%%%%%%% APPENDICES %%%%%%%%%%%%%%%%%%%%%

%%%%%%%%%%%%%%%%%%%% REFERENCES %%%%%%%%%%%%%%%%%%

\bibliographystyle{mnras}
\bibliography{vsi} 

\begin{thebibliography}{}
\makeatletter
\relax
\def\mn@urlcharsother{\let\do\@makeother \do\$\do\&\do\#\do\^\do\_\do\%\do\~}
\def\mn@doi{\begingroup\mn@urlcharsother \@ifnextchar [ {\mn@doi@}
  {\mn@doi@[]}}
\def\mn@doi@[#1]#2{\def\@tempa{#1}\ifx\@tempa\@empty \href
  {http://dx.doi.org/#2} {doi:#2}\else \href {http://dx.doi.org/#2} {#1}\fi
  \endgroup}
\def\mn@eprint#1#2{\mn@eprint@#1:#2::\@nil}
\def\mn@eprint@arXiv#1{\href {http://arxiv.org/abs/#1} {{\tt arXiv:#1}}}
\def\mn@eprint@dblp#1{\href {http://dblp.uni-trier.de/rec/bibtex/#1.xml}
  {dblp:#1}}
\def\mn@eprint@#1:#2:#3:#4\@nil{\def\@tempa {#1}\def\@tempb {#2}\def\@tempc
  {#3}\ifx \@tempc \@empty \let \@tempc \@tempb \let \@tempb \@tempa \fi \ifx
  \@tempb \@empty \def\@tempb {arXiv}\fi \@ifundefined
  {mn@eprint@\@tempb}{\@tempb:\@tempc}{\expandafter \expandafter \csname
  mn@eprint@\@tempb\endcsname \expandafter{\@tempc}}}

\bibitem[\protect\citeauthoryear{{Bai}}{{Bai}}{2017}]{bai17}
{Bai} X.-N.,  2017, \mn@doi [\apj] {10.3847/1538-4357/aa7dda}, \href
  {https://ui.adsabs.harvard.edu/abs/2017ApJ...845...75B} {845, 75}

\bibitem[\protect\citeauthoryear{{Bai} \& {Stone}}{{Bai} \&
  {Stone}}{2013}]{bs13}
{Bai} X.-N.,  {Stone} J.~M.,  2013, \mn@doi [\apj]
  {10.1088/0004-637X/769/1/76}, \href
  {https://ui.adsabs.harvard.edu/abs/2013ApJ...769...76B} {769, 76}

\bibitem[\protect\citeauthoryear{{Balbus} \& {Hawley}}{{Balbus} \&
  {Hawley}}{1991}]{bh91}
{Balbus} S.~A.,  {Hawley} J.~F.,  1991, \mn@doi [\apj] {10.1086/170270}, \href
  {https://ui.adsabs.harvard.edu/abs/1991ApJ...376..214B} {376, 214}

\bibitem[\protect\citeauthoryear{{Barker} \& {Latter}}{{Barker} \&
  {Latter}}{2015}]{bl15}
{Barker} A.~J.,  {Latter} H.~N.,  2015, \mn@doi [\mnras]
  {10.1093/mnras/stv640}, \href
  {https://ui.adsabs.harvard.edu/abs/2015MNRAS.450...21B} {450, 21}

\bibitem[\protect\citeauthoryear{{Barker} \& {Lithwick}}{{Barker} \&
  {Lithwick}}{2013}]{bl13}
{Barker} A.~J.,  {Lithwick} Y.,  2013, \mn@doi [\mnras]
  {10.1093/mnras/stt1561}, \href
  {https://ui.adsabs.harvard.edu/abs/2013MNRAS.435.3614B} {435, 3614}

\bibitem[\protect\citeauthoryear{{Bordes}, {Moisy}, {Dauxois}  \&
  {Cortet}}{{Bordes} et~al.}{2012}]{bordes12}
{Bordes} G.,  {Moisy} F.,  {Dauxois} T.,   {Cortet} P.-P.,  2012, \mn@doi
  [Physics of Fluids] {10.1063/1.3675627}, \href
  {https://ui.adsabs.harvard.edu/abs/2012PhFl...24a4105B} {24, 014105}

\bibitem[\protect\citeauthoryear{{Cassen}}{{Cassen}}{1993}]{cassen93}
{Cassen} P.,  1993, in Lunar and Planetary Science Conference. Lunar and
  Planetary Science Conference.
p.~261

\bibitem[\protect\citeauthoryear{{Chandrasekhar}}{{Chandrasekhar}}{1961}]{c61}
{Chandrasekhar} S.,  1961, {Hydrodynamic and hydromagnetic stability}

\bibitem[\protect\citeauthoryear{Craik}{Craik}{1986}]{craik86}
Craik A. D.~D.,  1986, Wave Interactions and Fluid Flows.
Cambridge Monographs on Mechanics, Cambridge University Press,
  \mn@doi{10.1017/CBO9780511569548}

\bibitem[\protect\citeauthoryear{{Cui} \& {Bai}}{{Cui} \& {Bai}}{2020}]{cb20}
{Cui} C.,  {Bai} X.-N.,  2020, \mn@doi [\apj] {10.3847/1538-4357/ab7194}, \href
  {https://ui.adsabs.harvard.edu/abs/2020ApJ...891...30C} {891, 30}

\bibitem[\protect\citeauthoryear{{Cui} \& {Bai}}{{Cui} \& {Bai}}{2021}]{cb21}
{Cui} C.,  {Bai} X.-N.,  2021, \mn@doi [\mnras] {10.1093/mnras/stab2220}, \href
  {https://ui.adsabs.harvard.edu/abs/2021MNRAS.507.1106C} {507, 1106}

\bibitem[\protect\citeauthoryear{{Cui} \& {Lin}}{{Cui} \& {Lin}}{2021}]{cl21}
{Cui} C.,  {Lin} M.-K.,  2021, \mn@doi [\mnras] {10.1093/mnras/stab1511}, \href
  {https://ui.adsabs.harvard.edu/abs/2021MNRAS.505.2983C} {505, 2983}

\bibitem[\protect\citeauthoryear{{D'Alessio}, {Cant{\"o}}, {Calvet}  \&
  {Lizano}}{{D'Alessio} et~al.}{1998}]{d98}
{D'Alessio} P.,  {Cant{\"o}} J.,  {Calvet} N.,   {Lizano} S.,  1998, \mn@doi
  [\apj] {10.1086/305702}, \href
  {https://ui.adsabs.harvard.edu/abs/1998ApJ...500..411D} {500, 411}

\bibitem[\protect\citeauthoryear{{Drazin}}{{Drazin}}{1977}]{drazin77}
{Drazin} P.~G.,  1977, \mn@doi [Proceedings of the Royal Society of London
  Series A] {10.1098/rspa.1977.0142}, \href
  {https://ui.adsabs.harvard.edu/abs/1977RSPSA.356..411D} {356, 411}

\bibitem[\protect\citeauthoryear{{Flaherty} et~al.,}{{Flaherty}
  et~al.}{2017}]{flaherty_etal17}
{Flaherty} K.~M.,  et~al., 2017, \mn@doi [\apj] {10.3847/1538-4357/aa79f9},
  \href {https://ui.adsabs.harvard.edu/abs/2017ApJ...843..150F} {843, 150}

\bibitem[\protect\citeauthoryear{{Flaherty}, {Hughes}, {Teague}, {Simon},
  {Andrews}  \& {Wilner}}{{Flaherty} et~al.}{2018}]{flaherty_etal18}
{Flaherty} K.~M.,  {Hughes} A.~M.,  {Teague} R.,  {Simon} J.~B.,  {Andrews}
  S.~M.,   {Wilner} D.~J.,  2018, \mn@doi [\apj] {10.3847/1538-4357/aab615},
  \href {https://ui.adsabs.harvard.edu/abs/2018ApJ...856..117F} {856, 117}

\bibitem[\protect\citeauthoryear{{Flaherty} et~al.,}{{Flaherty}
  et~al.}{2020}]{flaherty_etal20}
{Flaherty} K.,  et~al., 2020, \mn@doi [\apj] {10.3847/1538-4357/ab8cc5}, \href
  {https://ui.adsabs.harvard.edu/abs/2020ApJ...895..109F} {895, 109}

\bibitem[\protect\citeauthoryear{{Flock}, {Nelson}, {Turner}, {Bertrang},
  {Carrasco-Gonz{\'a}lez}, {Henning}, {Lyra}  \& {Teague}}{{Flock}
  et~al.}{2017}]{flock_etal17}
{Flock} M.,  {Nelson} R.~P.,  {Turner} N.~J.,  {Bertrang} G. H.~M.,
  {Carrasco-Gonz{\'a}lez} C.,  {Henning} T.,  {Lyra} W.,   {Teague} R.,  2017,
  \mn@doi [\apj] {10.3847/1538-4357/aa943f}, \href
  {https://ui.adsabs.harvard.edu/abs/2017ApJ...850..131F} {850, 131}

\bibitem[\protect\citeauthoryear{{Flock}, {Turner}, {Nelson}, {Lyra}, {Manger}
  \& {Klahr}}{{Flock} et~al.}{2020}]{flock_etal20}
{Flock} M.,  {Turner} N.~J.,  {Nelson} R.~P.,  {Lyra} W.,  {Manger} N.,
  {Klahr} H.,  2020, \mn@doi [\apj] {10.3847/1538-4357/ab9641}, \href
  {https://ui.adsabs.harvard.edu/abs/2020ApJ...897..155F} {897, 155}

\bibitem[\protect\citeauthoryear{{Flores-Rivera}, {Flock}  \&
  {Nakatani}}{{Flores-Rivera} et~al.}{2020}]{flores_etal20}
{Flores-Rivera} L.,  {Flock} M.,   {Nakatani} R.,  2020, \mn@doi [\aap]
  {10.1051/0004-6361/202039294}, \href
  {https://ui.adsabs.harvard.edu/abs/2020A&A...644A..50F} {644, A50}

\bibitem[\protect\citeauthoryear{{Fricke}}{{Fricke}}{1968}]{fricke68}
{Fricke} K.,  1968, \zap, \href
  {https://ui.adsabs.harvard.edu/abs/1968ZA.....68..317F} {68, 317}

\bibitem[\protect\citeauthoryear{{Fromang} \& {Lesur}}{{Fromang} \&
  {Lesur}}{2019}]{fl19}
{Fromang} S.,  {Lesur} G.,  2019, in EAS Publications Series. pp 391--413,
  \mn@doi{10.1051/eas/1982035}

\bibitem[\protect\citeauthoryear{{Galtier}, {Nazarenko}, {Newell}  \&
  {Pouquet}}{{Galtier} et~al.}{2003}]{galtier03}
{Galtier} S.,  {Nazarenko} S.~V.,  {Newell} A.~C.,   {Pouquet} A.,  2003, in
  {Velli} M.,  {Bruno} R.,  {Malara} F.,   {Bucci} B.,  eds,  American
  Institute of Physics Conference Series Vol. 679, Solar Wind Ten. pp 518--521,
  \mn@doi{10.1063/1.1618648}

\bibitem[\protect\citeauthoryear{{Gammie}, {Goodman}  \& {Ogilvie}}{{Gammie}
  et~al.}{2000}]{gammie00}
{Gammie} C.~F.,  {Goodman} J.,   {Ogilvie} G.~I.,  2000, \mn@doi [\mnras]
  {10.1046/j.1365-8711.2000.03669.x}, \href
  {https://ui.adsabs.harvard.edu/abs/2000MNRAS.318.1005G} {318, 1005}

\bibitem[\protect\citeauthoryear{{Goldreich} \& {Schubert}}{{Goldreich} \&
  {Schubert}}{1967}]{gs67}
{Goldreich} P.,  {Schubert} G.,  1967, \mn@doi [\apj] {10.1086/149360}, \href
  {https://ui.adsabs.harvard.edu/abs/1967ApJ...150..571G} {150, 571}

\bibitem[\protect\citeauthoryear{{Gressel}, {Turner}, {Nelson}  \&
  {McNally}}{{Gressel} et~al.}{2015}]{gressel_etal15}
{Gressel} O.,  {Turner} N.~J.,  {Nelson} R.~P.,   {McNally} C.~P.,  2015,
  \mn@doi [\apj] {10.1088/0004-637X/801/2/84}, \href
  {https://ui.adsabs.harvard.edu/abs/2015ApJ...801...84G} {801, 84}

\bibitem[\protect\citeauthoryear{{Gressel}, {Ramsey}, {Brinch}, {Nelson},
  {Turner}  \& {Bruderer}}{{Gressel} et~al.}{2020}]{gressel_etal20}
{Gressel} O.,  {Ramsey} J.~P.,  {Brinch} C.,  {Nelson} R.~P.,  {Turner} N.~J.,
   {Bruderer} S.,  2020, \mn@doi [\apj] {10.3847/1538-4357/ab91b7}, \href
  {https://ui.adsabs.harvard.edu/abs/2020ApJ...896..126G} {896, 126}

\bibitem[\protect\citeauthoryear{Kerswell}{Kerswell}{1999}]{K99}
Kerswell R.~R.,  1999, \mn@doi [Journal of Fluid Mechanics]
  {10.1017/S0022112098003954}, 382, 283–306

\bibitem[\protect\citeauthoryear{{Korycansky} \& {Pringle}}{{Korycansky} \&
  {Pringle}}{1995}]{kp95}
{Korycansky} D.~G.,  {Pringle} J.~E.,  1995, \mn@doi [\mnras]
  {10.1093/mnras/272.3.618}, \href
  {https://ui.adsabs.harvard.edu/abs/1995MNRAS.272..618K} {272, 618}

\bibitem[\protect\citeauthoryear{{Latter}}{{Latter}}{2016}]{latter16}
{Latter} H.~N.,  2016, \mn@doi [\mnras] {10.1093/mnras/stv2449}, \href
  {https://ui.adsabs.harvard.edu/abs/2016MNRAS.455.2608L} {455, 2608}

\bibitem[\protect\citeauthoryear{{Latter} \& {Papaloizou}}{{Latter} \&
  {Papaloizou}}{2017}]{lp17}
{Latter} H.~N.,  {Papaloizou} J.,  2017, \mn@doi [\mnras]
  {10.1093/mnras/stx2038}, \href
  {https://ui.adsabs.harvard.edu/abs/2017MNRAS.472.1432L} {472, 1432}

\bibitem[\protect\citeauthoryear{{Latter} \& {Papaloizou}}{{Latter} \&
  {Papaloizou}}{2018}]{lp18}
{Latter} H.~N.,  {Papaloizou} J.,  2018, \mn@doi [\mnras]
  {10.1093/mnras/stx3031}, \href
  {https://ui.adsabs.harvard.edu/abs/2018MNRAS.474.3110L} {474, 3110}

\bibitem[\protect\citeauthoryear{{Le Reun}, {Gallet}, {Favier}  \& {Le
  Bars}}{{Le Reun} et~al.}{2020}]{reun2020}
{Le Reun} T.,  {Gallet} B.,  {Favier} B.,   {Le Bars} M.,  2020, \mn@doi
  [Journal of Fluid Mechanics] {10.1017/jfm.2020.454}, \href
  {https://ui.adsabs.harvard.edu/abs/2020JFM...900R...2L} {900, R2}

\bibitem[\protect\citeauthoryear{{Lehmann} \& {Lin}}{{Lehmann} \&
  {Lin}}{2021}]{ll21}
{Lehmann} M.,  {Lin} M.-K.,  2021, arXiv e-prints, \href
  {https://ui.adsabs.harvard.edu/abs/2021arXiv211206153L} {p. arXiv:2112.06153}

\bibitem[\protect\citeauthoryear{{Lesur}}{{Lesur}}{2020}]{lesur20}
{Lesur} G.,  2020, arXiv e-prints, \href
  {https://ui.adsabs.harvard.edu/abs/2020arXiv200715967L} {p. arXiv:2007.15967}

\bibitem[\protect\citeauthoryear{{Lin}}{{Lin}}{2019}]{lin19}
{Lin} M.-K.,  2019, \mn@doi [\mnras] {10.1093/mnras/stz701}, \href
  {https://ui.adsabs.harvard.edu/abs/2019MNRAS.485.5221L} {485, 5221}

\bibitem[\protect\citeauthoryear{{Lin} \& {Youdin}}{{Lin} \&
  {Youdin}}{2015}]{ly15}
{Lin} M.-K.,  {Youdin} A.~N.,  2015, \mn@doi [\apj]
  {10.1088/0004-637X/811/1/17}, \href
  {https://ui.adsabs.harvard.edu/abs/2015ApJ...811...17L} {811, 17}

\bibitem[\protect\citeauthoryear{{Lubow} \& {Pringle}}{{Lubow} \&
  {Pringle}}{1993}]{lp93}
{Lubow} S.~H.,  {Pringle} J.~E.,  1993, \mn@doi [\apj] {10.1086/172669}, \href
  {https://ui.adsabs.harvard.edu/abs/1993ApJ...409..360L} {409, 360}

\bibitem[\protect\citeauthoryear{{Lyra} \& {Umurhan}}{{Lyra} \&
  {Umurhan}}{2019}]{lu19}
{Lyra} W.,  {Umurhan} O.~M.,  2019, \mn@doi [\pasp] {10.1088/1538-3873/aaf5ff},
  \href {https://ui.adsabs.harvard.edu/abs/2019PASP..131g2001L} {131, 072001}

\bibitem[\protect\citeauthoryear{{Malygin}, {Klahr}, {Semenov}, {Henning}  \&
  {Dullemond}}{{Malygin} et~al.}{2017}]{malygin_etal17}
{Malygin} M.~G.,  {Klahr} H.,  {Semenov} D.,  {Henning} T.,   {Dullemond}
  C.~P.,  2017, \mn@doi [\aap] {10.1051/0004-6361/201629933}, \href
  {https://ui.adsabs.harvard.edu/abs/2017A&A...605A..30M} {605, A30}

\bibitem[\protect\citeauthoryear{{Manger}, {Klahr}, {Kley}  \&
  {Flock}}{{Manger} et~al.}{2020}]{manger_etal20}
{Manger} N.,  {Klahr} H.,  {Kley} W.,   {Flock} M.,  2020, arXiv e-prints,
  \href {https://ui.adsabs.harvard.edu/abs/2020arXiv200809006M} {p.
  arXiv:2008.09006}

\bibitem[\protect\citeauthoryear{{Mied}}{{Mied}}{1976}]{mied76}
{Mied} R.~P.,  1976, \mn@doi [Journal of Fluid Mechanics]
  {10.1017/S0022112076002735}, \href
  {https://ui.adsabs.harvard.edu/abs/1976JFM....78..763M} {78, 763}

\bibitem[\protect\citeauthoryear{{Nazarenko} \& {Schekochihin}}{{Nazarenko} \&
  {Schekochihin}}{2011}]{ns11}
{Nazarenko} S.~V.,  {Schekochihin} A.~A.,  2011, \mn@doi [Journal of Fluid
  Mechanics] {10.1017/S002211201100067X}, \href
  {https://ui.adsabs.harvard.edu/abs/2011JFM...677..134N} {677, 134}

\bibitem[\protect\citeauthoryear{{Nelson}, {Gressel}  \& {Umurhan}}{{Nelson}
  et~al.}{2013}]{nelson_etal13}
{Nelson} R.~P.,  {Gressel} O.,   {Umurhan} O.~M.,  2013, \mn@doi [\mnras]
  {10.1093/mnras/stt1475}, \href
  {https://ui.adsabs.harvard.edu/abs/2013MNRAS.435.2610N} {435, 2610}

\bibitem[\protect\citeauthoryear{{Pfeil} \& {Klahr}}{{Pfeil} \&
  {Klahr}}{2019}]{pk19}
{Pfeil} T.,  {Klahr} H.,  2019, \mn@doi [\apj] {10.3847/1538-4357/aaf962},
  \href {https://ui.adsabs.harvard.edu/abs/2019ApJ...871..150P} {871, 150}

\bibitem[\protect\citeauthoryear{{Rafikov}}{{Rafikov}}{2007}]{rafikov07}
{Rafikov} R.~R.,  2007, \mn@doi [\apj] {10.1086/517599}, \href
  {https://ui.adsabs.harvard.edu/abs/2007ApJ...662..642R} {662, 642}

\bibitem[\protect\citeauthoryear{{Richard}, {Nelson}  \& {Umurhan}}{{Richard}
  et~al.}{2016}]{richard_etal16}
{Richard} S.,  {Nelson} R.~P.,   {Umurhan} O.~M.,  2016, \mn@doi [\mnras]
  {10.1093/mnras/stv2898}, \href
  {https://ui.adsabs.harvard.edu/abs/2016MNRAS.456.3571R} {456, 3571}

\bibitem[\protect\citeauthoryear{{Sch{\"a}fer}, {Johansen}  \&
  {Banerjee}}{{Sch{\"a}fer} et~al.}{2020}]{schafer_etal20}
{Sch{\"a}fer} U.,  {Johansen} A.,   {Banerjee} R.,  2020, \mn@doi [\aap]
  {10.1051/0004-6361/201937371}, \href
  {https://ui.adsabs.harvard.edu/abs/2020A&A...635A.190S} {635, A190}

\bibitem[\protect\citeauthoryear{Smith \& Waleffe}{Smith \&
  Waleffe}{1999}]{SW99}
Smith L.~M.,  Waleffe F.,  1999, \mn@doi [Physics of Fluids]
  {10.1063/1.870022}, 11, 1608

\bibitem[\protect\citeauthoryear{{Stoll} \& {Kley}}{{Stoll} \&
  {Kley}}{2014}]{sk14}
{Stoll} M. H.~R.,  {Kley} W.,  2014, \mn@doi [\aap]
  {10.1051/0004-6361/201424114}, \href
  {https://ui.adsabs.harvard.edu/abs/2014A&A...572A..77S} {572, A77}

\bibitem[\protect\citeauthoryear{{Teague} et~al.,}{{Teague}
  et~al.}{2016}]{teague_etal16}
{Teague} R.,  et~al., 2016, \mn@doi [\aap] {10.1051/0004-6361/201628550}, \href
  {https://ui.adsabs.harvard.edu/abs/2016A&A...592A..49T} {592, A49}

\bibitem[\protect\citeauthoryear{{Teed} \& {Latter}}{{Teed} \&
  {Latter}}{2021}]{tl21}
{Teed} R.~J.,  {Latter} H.~N.,  2021, \mn@doi [\mnras]
  {10.1093/mnras/stab2311}, \href
  {https://ui.adsabs.harvard.edu/abs/2021MNRAS.507.5523T} {507, 5523}

\bibitem[\protect\citeauthoryear{{Turner}, {Fromang}, {Gammie}, {Klahr},
  {Lesur}, {Wardle}  \& {Bai}}{{Turner} et~al.}{2014}]{turner_etal14}
{Turner} N.~J.,  {Fromang} S.,  {Gammie} C.,  {Klahr} H.,  {Lesur} G.,
  {Wardle} M.,   {Bai} X.~N.,  2014, in {Beuther} H.,  {Klessen} R.~S.,
  {Dullemond} C.~P.,   {Henning} T.,  eds, Protostars and Planets VI. p.~411
  (\mn@eprint {arXiv} {1401.7306}),
  \mn@doi{10.2458/azu_uapress_9780816531240-ch018}

\bibitem[\protect\citeauthoryear{{Urpin} \& {Brandenburg}}{{Urpin} \&
  {Brandenburg}}{1998}]{ub98}
{Urpin} V.,  {Brandenburg} A.,  1998, \mn@doi [\mnras]
  {10.1046/j.1365-8711.1998.01118.x}, \href
  {https://ui.adsabs.harvard.edu/abs/1998MNRAS.294..399U} {294, 399}

\bibitem[\protect\citeauthoryear{Wersinger, Finn  \& Ott}{Wersinger
  et~al.}{1980}]{Wersinger+80}
Wersinger J.,  Finn J.~M.,   Ott E.,  1980, \mn@doi [The Physics of Fluids]
  {10.1063/1.863116}, 23, 1142

\makeatother
\end{thebibliography}

%%%%%%%%%%%%%%%%%%%%%%%%%%%%%%%%%%%%%%%%%%%%%%%%%%

\bsp	% typesetting comment
\label{lastpage}
\end{document}